\documentclass[a4paper,usenatbib,onecolumn]{mnras}

\usepackage[T1]{fontenc}
\usepackage{ae,aecompl}
\newenvironment{sistema}%
{\left\lbrace\begin{array}{@{}l@{}}}%
{\end{array}\right.}
\usepackage{amssymb}
\usepackage{multirow}

\usepackage{graphicx}	
\usepackage{amsmath}	
\usepackage{amssymb}	
\usepackage{float}
\usepackage{tensor}	
\usepackage{soul}

\newcommand{\bear}{\begin{eqnarray}}

\newcommand{\eear}{\end{eqnarray}}

\newlength{\tskip}
\newbox\pippobox

\def\be{\begin{equation}}
\def\ee{\end{equation}}
\def\bea{\begin{eqnarray}}
\def\eea{\end{eqnarray}}

\newbox\pippobox

\def\be{\begin{equation}}
\def\ee{\end{equation}}
\def\bea{\begin{eqnarray}}
\def\eea{\end{eqnarray}}

\def\9{\nabla}

\def\dd{{\rm d}}

\def\P{\Phi}

\def\nn{\nonumber}

\def\le{\left}
\def\ri{\right}
\def\6{\partial}

\def\f{\frac}

\def\la{\langle}

\def\0{(0)}

\def\>{\rightarrow}

\title[Testing Hu-Sawicki $f(R)$ gravity with the Effective Field Theory approach]{Testing Hu-Sawicki $f(R)$ gravity with the Effective Field Theory approach}

\author[B. Hu et al.]{
Bin Hu,$^{1,~2}$\thanks{E-mail: binhu@icc.ub.edu}
Marco Raveri,$^{3,~4,~5}$
Matteo Rizzato$^{6}$
and Alessandra Silvestri$^{1}$
\\
\\
$^{1}$Institute Lorentz, Leiden University, PO Box 9506, Leiden 2300 RA, The Netherlands\\
$^{2}$Institut de Ci{\`e}ncies del Cosmos (ICCUB), Universitat de Barcelona (IEEC-UB), Mart{\'\i} i Franqu{\`e}s 1, E08028 Barcelona, Spain\\
$^{3}$SISSA - International School for Advanced Studies, Via Bonomea 265, 34136, Trieste, Italy \\
$^{4}$INFN, Sezione di Trieste, Via Valerio 2, I-34127 Trieste, Italy \\
$^{5}$INAF-Osservatorio Astronomico di Trieste, Via G.B. Tiepolo 11, I-34131 Trieste, Italy \\
$^{6}$Dipartimento di Fisica e Astronomia ``G. Galilei'', \break Universit\`a degli Studi di Padova, via Marzolo 8, I-35131, Padova, Italy
}

\date{Accepted XXX. Received YYY; in original form ZZZ}
\pubyear{2015}

\begin{document}
\label{firstpage}
\pagerange{\pageref{firstpage}--\pageref{lastpage}}
\maketitle

\begin{abstract}
We show how to fully map a specific  model of modified gravity into the  Einstein-Boltzmann solver EFTCAMB.
This approach consists in few steps and allows to obtain the cosmological phenomenology of a model with minimal effort.
We discuss all these steps, from the solution of the dynamical equations for the cosmological background of the model to the use of the mapping relations to cast the model into the effective field theory language and use the latter to solve for  perturbations.
We choose the Hu-Sawicki $f(R)$ model of gravity as our working example.
After solving the background and performing the mapping, we interface the algorithm with EFTCAMB and  take advantage of the effective field theory framework to integrate the full dynamics of linear perturbations,  returning all quantities needed to accurately compare the model with observations.
We discuss some observational signatures of this model, focusing on the linear growth of cosmic structures. In particular we present the behavior of $f\sigma_8$ and $E_G$ that, unlike the $\Lambda$CDM scenario, are generally scale dependent in addition to redshift dependent.
Finally, we study the observational implications of the model by comparing its cosmological predictions to the Planck 2015 data, including CMB lensing, the WiggleZ galaxy survey and the CFHTLenS weak lensing survey measurements.
We find that while WiggleZ data favor a non-vanishing value of the Hu-Sawicki model parameter, $\log_{10}(-f^0_{R})$, and consequently a large value of $\sigma_8$,  CFHTLenS drags the estimate of $\log_{10}(-f^0_{R})$ back to the $\Lambda$CDM limit.
\end{abstract}




\section{Introduction}\label{Sec:Intro}

Cosmic acceleration still remains an open challenge for modern cosmology. 
As we gear up for highly accurate upcoming and future data from cosmological surveys, it is becoming increasingly important to equip ourselves with a unified and accurate framework to compare models of dark energy and modified gravity (DE/MG) with observational data.  \\
To this extent a very promising approach is represented by the effective field theory (EFT) of cosmic acceleration, introduced in~\cite{Gubitosi:2012hu,Bloomfield:2012ff}, inspired by EFT of inflation and large scale structure~\cite{Creminelli:2006xe,Cheung:2007st,Weinberg:2008hq,Creminelli:2008wc,Park:2010cw,Jimenez:2011nn,Carrasco:2012cv,Hertzberg:2012qn,Carrasco:2013mua,Porto:2013qua,Senatore:2014vja}. 
Being based on a parametrized action, it offers a unifying and model-independent framework to perform agnostic tests of gravity as well as to accurately explore most of the viable models of cosmic acceleration. 
Indeed, any model of DE/MG which introduces one additional scalar degree of freedom and allows for a well defined Jordan frame, can be mapped exactly into the EFT language, without the need of resorting to any approximation. \\
In a further leap towards observations, this framework has been implemented into the Einstein-Boltzmann solver\\ 
CAMB/CosmoMC~\footnote{\url{http://camb.info}}~\cite{Lewis:1999bs,Lewis:2002ah},  resulting in the powerful patches EFTCAMB/EFTCosmoMC, introduced in~\cite{Hu:2013twa,Raveri:2014cka,Hu:2014oga,Hu:2014sea}, which are publicly available at~\url{http://wwwhome.lorentz.leidenuniv.nl/~hu/codes/}. 
The latter fully exploit the unifying nature of the EFT formalism, and provide a highly accurate and efficient setup with which to test gravity on cosmological scales. Model independent parametrizations as well as specific models of DE/MG can be explored with the same code, without the need of specializing the set of perturbed equations to the case under study. These equations are indeed implemented in the EFT language, and they are fully evolved by EFTCAMB, without the use of any quasi-static approximation. All that is needed is to  define or map the chosen parametrization or model into the EFT formalism. In particular, when exploring specific models of modified gravity, within the so called \emph{mapping} approach, there are  two possibilities: the designer approach, in which a class of models is chosen but their functional form is reconstructed from requiring a given expansion history; the \emph{full mapping} approach, in which a model is fully specified and one needs to solve for its background dynamics before translating it into the EFT formalism. After the background is worked out with any of these two approaches, EFTCAMB takes on and evolves the full set of perturbation equations, without the need from the user to calculate these equations in a model specific way. Several models of DE/MG are already implemented in the publicly available version of EFTCAMB, including $f(R)$ gravity. However, most of them use the designer approach, except for the recently added low-energy Ho\v rava gravity~\cite{Frusciante:2015maa} which, however, did not require a separate treatment of the background.

In this paper we present the first full implementation of a model of modified gravity, illustrating all the steps from the construction of a model-specific background solver, the mapping into the EFT formalsim and the  implementation of the solver into EFTCAMB. The model that we choose for this purpose is the popular Hu-Sawicki  $f(R)$ gravity, that was introduced in~\cite{Hu:2007nk} and represents one of the few known viable functional forms of $f(R)$ with  the interesting feature of being able to satisfy solar system tests of gravity. The non-linear structure formation via N-body simulation of this model are studied in \cite{Zhao:2010qy,Baldi:2013iza,Llinares:2013jza,Lombriser:2013eza,Winther:2015wla}. In this paper we will focus on its linear perturbation phenomena and use it as a way to illustrate the exact implementation of specific models of DE/MG into the EFT framework and its corresponding Einstein-Boltzmann solver EFTCAMB. 

\section{The EFT framework and EFTCAMB}\label{Sec:Map}
The effective field theory (EFT) approach to dark energy/modified gravity (DE/MG) was first proposed in~\cite{Gubitosi:2012hu,Bloomfield:2012ff}, and further investigated and developed in~\cite{Gleyzes:2013ooa,Bloomfield:2013efa,Piazza:2013coa,Gleyzes:2014rba,Gleyzes:2014dya}. Focusing on linear cosmological perturbations around a Friedmann-Lemaitre-Robertson-Walker (FLRW) universe, it is based on a Jordan frame action built in unitary gauge out of  all operators which are invariant under time-dependent spatial diffeomorphisms and up to quadratic in perturbations. It offers a model-independent and unifying framework to study  all  viable single scalar field theories of DE/MG which allow a well defined Jordan frame, such as $f(R)$ gravity, quintessence and, more generally, the Horndeski class of theories and beyond~\cite{Frusciante:2015maa,Frusciante:2016xoj}  (see~\cite{Clifton:2011jh,Bull:2015stt} for a review of models of MG/DE). In this approach, the additional scalar degree of freedom representing DE/MG is eaten by the metric via a foliation of space-time into  constant time hyper-surfaces that correspond to uniform scalar field ones. Up to the quadratic order in perturbations, the action reads
\begin{align}
\mathcal{S}_{\rm EFT} = \int d^4x &\sqrt{-g}  \bigg\{ \frac{m_0^2}{2} \left[1+\Omega(\tau)\right] R + \Lambda(\tau) - c(\tau)\,a^2\delta g^{00} + \frac{M_2^4 (\tau)}{2} \left( a^2\delta g^{00} \right)^2
 - \frac{\bar{M}_1^3 (\tau)}{2} \, a^2\delta g^{00}\,\delta \tensor{K}{^\mu_\mu}  
    - \frac{\bar{M}_2^2 (\tau)}{2} \left( \delta \tensor{K}{^\mu_\mu}\right)^2 \nn \\
   & - \frac{\bar{M}_3^2 (\tau)}{2} \,\delta \tensor{K}{^\mu_\nu}\,\delta \tensor{K}{^\nu_\mu}
      + m_2^2(\tau)\left(g^{\mu\nu}+n^{\mu} n^{\nu}\right)\partial_{\mu}(a^2g^{00})\partial_{\nu}(a^2g^{00}) +\frac{\hat{M}^2(\tau)}{2} \, a^2 \delta g^{00}\,\delta \mathcal{R}+	\ldots \bigg\} + S_{m} [g_{\mu \nu}, \chi_m ],\label{actioneft}
\end{align}
where $R$ is the four-dimensional Ricci scalar, $\delta g^{00}$, $\delta \tensor{K}{^\mu_\nu}$, $\delta \tensor{K}{^\mu_\mu}$ and  $\delta \mathcal{R}$ are respectively the perturbations of the upper time-time component of the metric, the extrinsic curvature and its trace and the  three dimensional spatial Ricci scalar of the constant-time hypersurfaces. Finally,  $S_m$ is the matter action. 
Since the choice of the unitary gauge breaks time diffeomorphism invariance, each operator in the action can be multiplied by a time-dependent coefficient; in our convention, $\{\Omega,\Lambda,c, M_2^4,\bar{M}_1^3,\bar{M}_2^2,\bar{M}_2^2,\bar{M}_3^2,m_2^2,\hat{M}^2\}$ are unknown functions of the conformal time, $\tau$, and we will refer to them as EFT functions.
Only three of these functions, namely $\{\Omega,c,\Lambda\}$, affect the dynamics of the background and we wil refer to them as background EFT functions. Of course, they contribute also to the dynamics of linear perturbations, along with the remaining functions. It is interesting to notice that, using the Friedmann equations, two of the background functions can be eliminated in favor of the Hubble parameter $H(z)$ and its derivative. In other words, two free functions of time are sufficient to fix the background dynamics. Different generalizations of action~(\ref{actioneft}) have been recently studied in the literature~\cite{} to include further DE/MG models. However, as we shall see, the first line of~(\ref{actioneft}) is all we need in order to study $f(R)$ models via the EFT formalism.

In the action Eq.(\ref{actioneft}), the  extra scalar degree of freedom is hidden inside the metric perturbations, however in order to study the dynamics of linear perturbations and  investigate the stability of a given model, it is convenient to make it explicit by means of  the St$\ddot{\text{u}}$ckelberg technique i.e. performing an infinitesimal coordinate transformation such that $\tau\rightarrow \tau+\pi$, where the new field $\pi$ is the St$\ddot{\text{u}}$ckelberg field which describes the extra propagating degree of freedom.
Varying the action with respect to the $\pi$-field one obtains a dynamical perturbative equation for the extra degree of freedom which allows to control directly the stability of the theory, as discussed in~\cite{Hu:2013twa}. 

In reference~\cite{Hu:2013twa,Raveri:2014cka} the EFT framework has been implemented into CAMB/CosmoMC creating the EFTCAMB/EFTCosmoMC patches which are publicly available (see~\cite{Hu:2014oga} for the implementation details).  EFTCAMB evolves the full equations for linear perturbations without relying on any quasi-static approximation. In addition to the standard matter components ({\it i.e.} dark matter, baryon, radiation and massless neutrinos), massive neutrinos have also been included~\cite{Hu:2014sea}.  

There are  two ways to treat the EFT functions, which correspond to the twofold nature of the EFT formalism and that are both implemented in EFTCAMB. 
In a first case, one can simply treat them, in a phenomenological way, as unknown functions and parametrize their dependence on time to agnostically explore the space of DE/MG models. 
We typically refer to this approach as the \emph{pure} EFT one.  
Alternatively, one can specialize to a given model of DE/MG and map its covariant formulation into the EFT language and use EFTCAMB to study the exact dynamics of perturbations, for that specific model (usually referred to as \emph{mapping} mode).
In the latter case one can treat the background via a designer approach, {\it i.e.} fixing the expansion history and reconstructing the specific model in terms of EFT functions; or in a full mapping approach, {\it i.e.} one can choose a fully specified model and  solve for its background.  \\
Furthermore, the code has a powerful built-in module that investigates whether a chosen model is  viable,  through a set of  general conditions of mathematical and physical stability. In particular,  the physical requirements include the avoidance of ghost and gradient instabilities for both the scalar and the tensor degrees of freedom. The stability requirements are translated into \emph{viability  priors} on the parameter space when using EFTCosmoMC to interface EFTCAMB with cosmological data, and they can sometimes dominate over the constraining power of data~\cite{Raveri:2014cka}. 

In this paper we focus on the \emph{full mapping} branch and consider a specific model of $f(R)$ gravity as an example, namely the Hu-Sawicki model~\cite{Hu:2007nk}.  We present in detail its full implementation into EFTCAMB, all the way from the  mapping of the model into the EFT language to the evolution of the full dynamics of perturbations.
There exists already a  publicly available linear Einstein-Boltzmann solver for $f(R)$ models, i.e.  FRCAMB\footnote{\url{http://darklight.brera.inaf.it/cosmonews/frcamb/}}. This package was first developed by~\cite{He:2012wq} for the designer $f(R)$ models, then extended by~\cite{Xu:2015usa} to include the Hu-Sawicki models. Here we would like to emphasize that for the study of linear perturbation phenomena in a specific models, the EFT approach gives not only the same results as the specific codes, but also it is able to treat in a unified way most of the viable single scalar field theories of DE/MG which have a well defined Jordan frame representation. This advantage makes EFTCAMB more robust and convenient to use for testing gravity. 

\section{Hu-Sawicki model}\label{theory}

The Hu-Sawicki model of $f(R)$ gravity was introduced in~\cite{Hu:2007nk} and represents one of the few known viable functional forms of $f(R)$ with  the interesting feature of being able to satisfy solar system tests of gravity. It corresponds to the following action:
\be\label{fRaction}
S=\int d^4x \sqrt{-g}  \frac{m_0^2}{2} \left[R +f(R)\right]+ S_{m} [g_{\mu \nu}, \chi_m ],
\ee
where
\be\label{Hu-Sawicki}
f(R)=-m^2\f{c_1(R/m^2)^n}{c_2(R/m^2)^n+1}\,\qquad\textrm{with}\qquad m^2\equiv H_0^2\Omega_m=(8315{\rm Mpc})^{-2}\le(\f{\Omega_m h^2}{0.13}\ri)\,;
\ee
The non-linear terms in $f(R)$ introduce higher-order derivatives acting on the metric, making explicit the higher order nature of the theory. It is possible to cast the theory into a second order one with an extra  scalar degree of freedom represented by $f_R\equiv\dd f/\dd R$, \emph{i.e.} the \emph{scalaron}.  

In the high curvature regime, $R\gg m^2$, the action~(\ref{Hu-Sawicki}) can be expanded in $m^2/R$:
\be\label{Hu-SawickiApp}
\lim_{m^2/R\>0}f(R)\approx -\f{c_1}{c_2}m^2+\f{c_1}{c_2^2}m^2\le(\f{m^2}{R}\ri)^n+\cdots\;.
\ee
This limit can be applied up to $z=0$ if the parameter $m^2$ is properly chosen. In particular it can be applied for the value given in (\ref{Hu-Sawicki}), as discussed in~\cite{Hu:2007nk}.
Furthermore one can notice that the first term corresponds to a cosmological constant  and the second term to a deviation from it, which becomes more important at low curvature. 
It is possible to closely mimic a $\Lambda$CDM evolution if the value of $c_1/c_2$ is fixed by:
\be\label{C1OverC2}
\f{c_1}{c_2}\approx 6\f{\Omega_{\Lambda}}{\Omega_m}\;.
\ee
which is valid as long as the high curvature regime holds. Using the latter relation, the number of free model parameters can be reduced to two.  
In addition, the parameter $c_2$ can be expressed in terms of $f_R^0 \equiv \dd f/\dd R(z=0)$ so that the two free parameters of the Hu-Sawicki model, that we shall discuss in the next sections, are $f_R^0$ and $n$.

\subsection{Mapping to the EFT framework}\label{Sec:Mapping}
\begin{figure}
\centering
\includegraphics[width=1\textwidth]{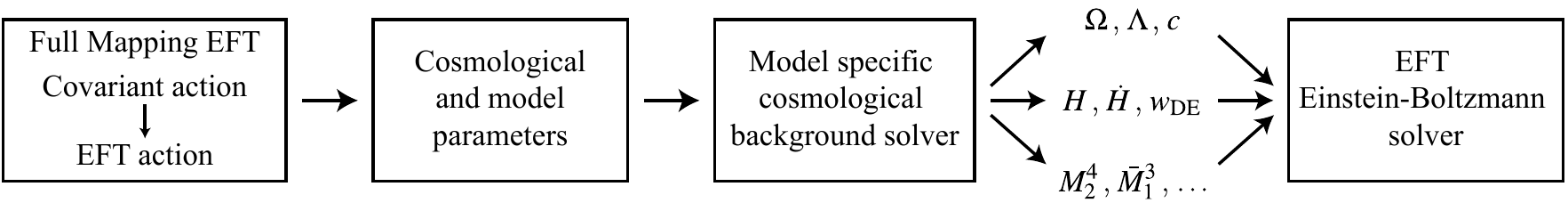}
\caption{Diagram illustrating the logical steps of the implementation of a full-mapping module in EFTCAMB. At first a specific theory needs to be mapped to the EFT framework. These mapping relations, along with the cosmological and model parameters are fed to a module that solves the cosmological background equations, for the specific theory, and outputs the time evolution of the EFT functions. These functions are then used to evolve the full perturbed Einstein-Boltzmann equations and compute cosmological observables.
}
\label{fig:fm_demo}
\end{figure}

It is  straightforward to map $f(R)$ models of gravity into the EFT formalism. In particular this has been already presented in~\cite{Gubitosi:2012hu,Bloomfield:2012ff} and here we will briefly summarize the main steps and the final result. Starting from action~(\ref{fRaction}), we can expand it in perturbations of the Ricci scalar around its value on a FLRW background, $R^{(0)}$. It turns out to be convenient to do so by choosing a preferred time-slicing for which the constant time hypersurfaces coincides with the uniform $R$ hypersurfaces. This allows to truncate the expansion at linear order since all higher order terms will contribute always at least one power of $\delta R$ to the equations, and the latter  vanishes.  Therefore, to linear order we have:
\be\label{fR_exp}
S=\int\,d^4x\sqrt{-g}\frac{m_0^2}{2}\left\{\left[1+f_R(R^{(0)})\right]R+f(R^{(0)})-R^{(0)}f_R(R^{(0)})\right\}\,.
\ee
Comparing this action with the EFT one~(\ref{actioneft}), we can easily derive the following mapping recipe:
\begin{align}\label{fR_matching}
\Lambda=\frac{m_0^2}{2}\left[f- Rf_R \right] \hspace{0.5cm};\hspace{0.5cm} c=0 \hspace{0.5cm};\hspace{0.5cm} \Omega=f_R \,.
\end{align}
As anticipated, the dynamics of linear perturbations in $f(R)$ models of gravity can be studied exactly through the EFT framework working only with two EFT functions: $\Omega$ and $\Lambda$.  In figure. \ref{fig:fm_demo}, we demonstrate the logical steps of the implementation of a full-mapping module in EFTCAMB. Our algorithms only asks for a few steps to interface the model parameters with EFTCAMB, such as the background equation of state of the dynamical dark energy field (or the modified Hubble parameter) and the corresponding EFT functions. The modules of the code will automatically calculate the derivatives and integrals of these quantities numerically. In the following we will show how to determine the time-dependence of these functions, which is all is needed in order to have EFTCAMB solve for the dynamics of cosmological perturbations, in $f(R)$ models, through the full mapping procedure. 

\subsection{Solving the cosmological background}\label{Sec:Background}

\begin{figure}
\centering
\includegraphics[width=\textwidth]{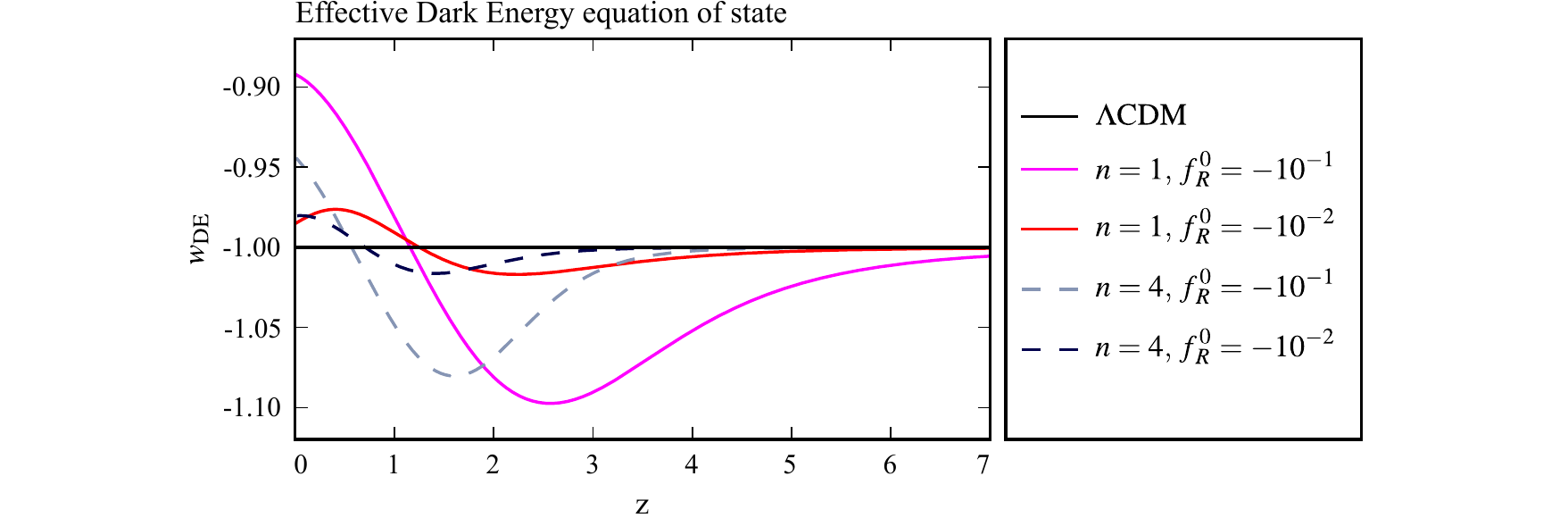}
\caption{Effective Dark Energy Equation of State for the Hu-Sawicki $f(R)$ model as a function of redshift. Different colors and ticks represent different values of model parameters, as shown in legend.}
\label{fig:eos}
\end{figure}

We shall now focus on solving the background dynamics of the Hu-Sawicki model, with the goal of deriving the time-dependence of the EFT functions in~(\ref{fR_matching}). Let us start from the modified Friedmann equations in conformal time:
\begin{equation}
 \label{friedmandmod}
 \begin{aligned}
 &\big{(} 1+f_R\big{)}\mathcal{H}^2 + a^2 \frac{f}{6} - \frac{\ddot{a}}{a}f_R + \mathcal{H}\dot{f}_R = \frac{1}{3m_0^2}a^2\rho_{m} \,,\\
 &\frac{\ddot{a}}{a} - \left( 1+f_R\right)\mathcal{H}^2 + \frac{a^2}{6}f + \frac{\ddot{f}_R}{2} = -\frac{1}{6m_0^2}a^2\left( \rho_{m} + 3P_{m}\right) \,.
 \end{aligned}
\end{equation}
The trace of  Einstein equations  provides a dynamical equation for the additional scalar degree of freedom of the theory
\begin{equation}\label{fR_eq}
\Box f_R = \frac{1}{3}\left( -\frac{\rho_{m} }{m_0^2}+ R - f_R R + 2f \right) \,.
\end{equation} 
From a closer look at~(\ref{fR_eq}), it can be realized that in the high curvature regime  the solution of this equation displays highly oscillatory modes when $f_{RR}>0$,~\cite{Song:2006ej,Starobinsky:2007hu}. These oscillations have an amplitude that decays in time and eventually fades away as the curvature decreases. Nevertheless, one needs extra care to cope with these oscillations when solving the background dynamics, as we will discuss later.

Starting from~(\ref{friedmandmod}) it is possible to recast the Friedmann equations in their general relativistic form, incorporating  the effects of the modifications of gravity into an effective dark fluid with the following equation of state:
\begin{align}\label{eff_fluid}
 w_{\rm DE} &=  -\frac{1}{3} - \frac{2}{3}\frac{\left(\mathcal{H}^2f_R - \frac{a^2f}{6} -\frac{1}{2}\ddot{f}_R\right)}{\left(-\mathcal{H}^2f_R - \frac{a^2f}{6} - \mathcal{H}\dot{f}_R +\frac{a^2f_R R}{6}\right)}\;.
\end{align}
Following~\cite{Hu:2007nk}, we can then define two auxiliary variables:
\begin{equation}
\label{yhyr}
y_h = \left(\frac{\mathcal{H}^2}{m^2} -a^{-1}\right) a^{-2},\quad y_r = \frac{R}{m^2} - 3a^{-3},
\end{equation}
to recast the first equation in~(\ref{friedmandmod}) and the geometrical relation between $R$ and $H$ into a first order system of non-linear ordinary differential equations:
\begin{equation}\label{new_systemtosolve}
 \begin{sistema}
  y_h^{\prime} = \frac{1}{3} y_r - 4 y_h\\
  \\
  y_r^{\prime} = 9a^{-3} - \frac{1}{y_h+a^{-3}}\frac{m^2}{f_{RR}}\Big{[} y_h - f_R\Big{(} \frac{1}{6}y_r - y_h - \frac{1}{2}a^{-3} \Big{)} + \frac{1}{6}\frac{f}{m^2}\Big{]}.
 \end{sistema}
\end{equation}
where the prime stands for the derivative w.r.t.  the number of efolds $\ln a$.
As we already discussed, the solution of this system of differential equations is expected to display high frequency oscillations at high redshift. In order to ensure an accurate and efficient approach, we shall determine the particular solution of it around which oscillations happen. 
This is defined by the smooth evolution of the minimum of the potential defined by the r.h.s. of equation~(\ref{fR_eq}). 
Requiring the field to seat at the bottom of  its potential, immediately results in:
\begin{equation}\label{Ricci_Scalar}
 R = \frac{k^2\rho_{m}-2f}{1-f_R}.
\end{equation}
We can then linearize the initial system around the evolution of the minimum of the effective potential.
At the leading order in the high curvature expansion~(\ref{Hu-SawickiApp}), the evolution for $\mathcal{H}^2$ can be described by:
\begin{equation}
\mathcal{H}^2 \approx  m^2a^{-1} + \frac{c_1}{6c_2}m^2a^2 \,.
\end{equation}
Then, the variable that can be safely used for our linearization at high redshift, is:
\begin{equation}\label{new_variable}
 \bar{y}_h \equiv \frac{\mathcal{H}^2 - m^2a^{-1} - \frac{c_1}{6c_2}m^2a^2}{m^2a^{-1}}  = \frac{\mathcal{H}^2 - \mathcal{H}^2_{\Lambda{\rm CDM}}}{m^2a^{-1}}\;,
\end{equation}
where $\mathcal{H}^2_{\Lambda{\rm CDM}}$ is the Hubble parameter for the $\Lambda$CDM model.
Once~(\ref{new_systemtosolve}) has been linearized, the evolution of its stable solution is well approximated by the ratio of the non-homogeneous term and the mass term of the corresponding linearized system.

The strategy that we shall use to solve the background equations, for the Hu-Sawicki $f(R)$ model, consists in the following steps:
we will start deep in radiation domination, when the system~(\ref{new_systemtosolve}) shows extreme stiffness, and we approximate its solution with the stable solution of the linearized system;
when the stiffness of the system becomes tractable by standard numerical methods the particular solution introduced before, is used as the initial condition for the numerical solution of~(\ref{new_systemtosolve}).
After solving the background equations, we can reconstruct the effective DE equation of state~(\ref{eff_fluid}):
\begin{equation}\label{EOS_DE}
 w_{\rm DE}=-1-\frac{y'_h}{3y_h} \,.
\end{equation}
Figure~\ref{fig:eos} shows the time evolution of this quantity as a function of redshift, for some selected values of model parameters. 
As we can see the Hu-Sawicki model is never exactly mimicking a cosmological constant. Differences, with respect to the $\Lambda$CDM expansion history, are however decreasing as we decrease the value of $f_{R}^0$ and we increase the value of the exponent $n$, as expected.

The numerical algorithm to solve~(\ref{new_systemtosolve}), along with the mapping~(\ref{fR_matching}), constitute all the ingredients that the user needs to supply in order to have EFTCAMB solve for the full dynamics of linear perturbations in the Hu-Sawicki model. All these operations can then be implemented in a model specific module of EFTCAMB that is simply interfaced with the part that solves Boltzmann-Einstein equations.

While we specialized to the Hu-Sawicki model as a working example, the steps presented in this section also constitute the necessary ingredients to map other modified gravity models into EFTCAMB. Indeed, in a similar way one has to write a model specific module that solves the background equations for the model of interest. Then one has to work out and use the mapping relations between the model and the EFT framework to get the specific time dependence of the EFT functions (e.g.~\cite{Frusciante:2016xoj} offers a complete guide on the mapping). Once these functions are fed to EFTCAMB, the code will automatically calculate derivatives and  integrals numerically and finally output all the cosmological observables of interest.

\section{Cosmological observables}\label{Sec:HSCosmology}
%

\begin{figure}
\centering
\includegraphics[width=1\textwidth]{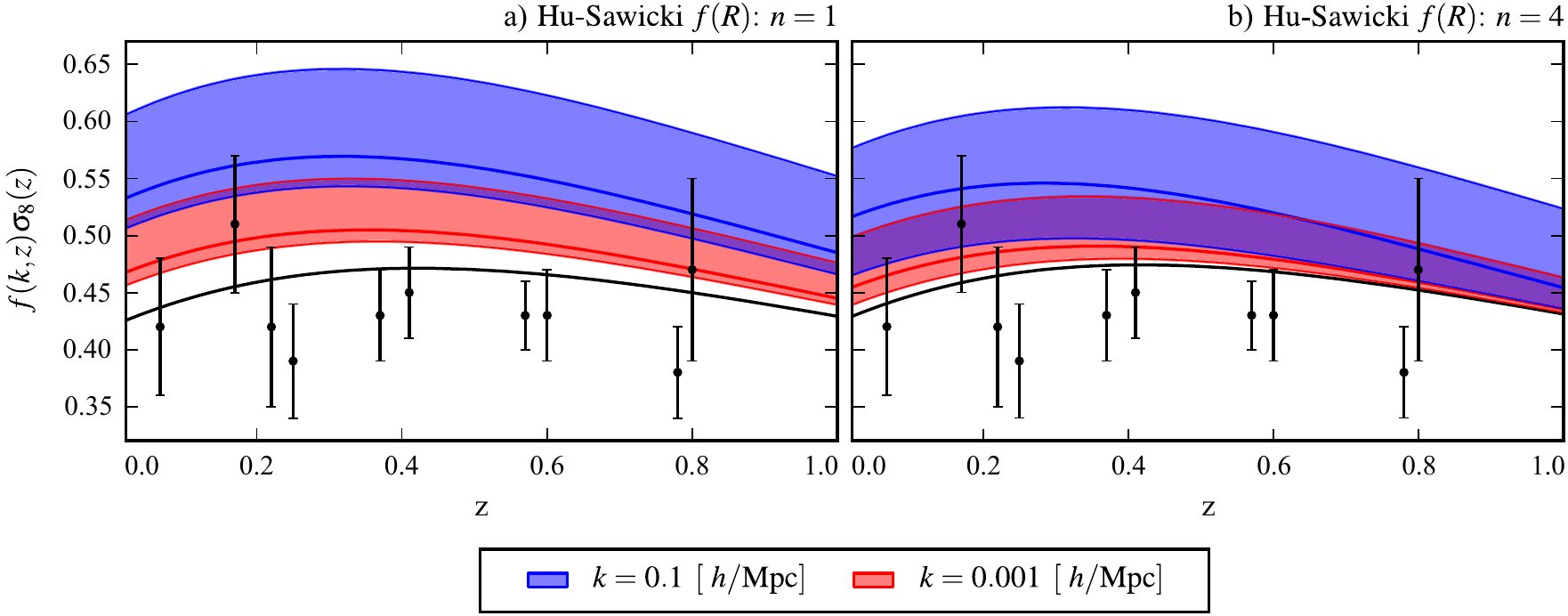}
\caption{The linear growth rate, $f\sigma_8$, in Hu-Sawicki $f(R)$ gravity, at different redshifts and scales. Different panels correspond to different values of the Hu-Sawicki exponent $n$. 
The black thick line corresponds to $\Lambda$CDM cosmology.
Different colors indicate different scales at which $f\sigma_8$ is computed, as shown in the legend.
The colored thick lines show the behavior of $f\sigma_8$ for the mean value of cosmological parameters of the corresponding model, as obtained from the most complete data set combination discussed in section~\ref{Sec:CosmologicalConstraints}.
The color bands denote the $1\sigma$ uncertainties on the $f^0_R$ parameter from the data compilation of Planck15+BAO+JLA+WiggleZ+CFHTLenS.
The black points with error bars are 10 redshift space distortion measurements from: 6dFGRS~\protect\cite{2012MNRAS.423.3430B}, 2dFGRS~\protect\cite{Percival:2004fs}, WiggleZ~\protect\cite{2011MNRAS.415.2876B}, SDSS LRG~\protect\cite{2012MNRAS.420.2102S}, BOSS CMASS~\protect\cite{2012MNRAS.426.2719R} and VIPERS~\protect\cite{delaTorre:2013rpa}.}
\label{fig:fs8}
\end{figure}

\begin{figure}
\centering
\includegraphics[width=\textwidth]{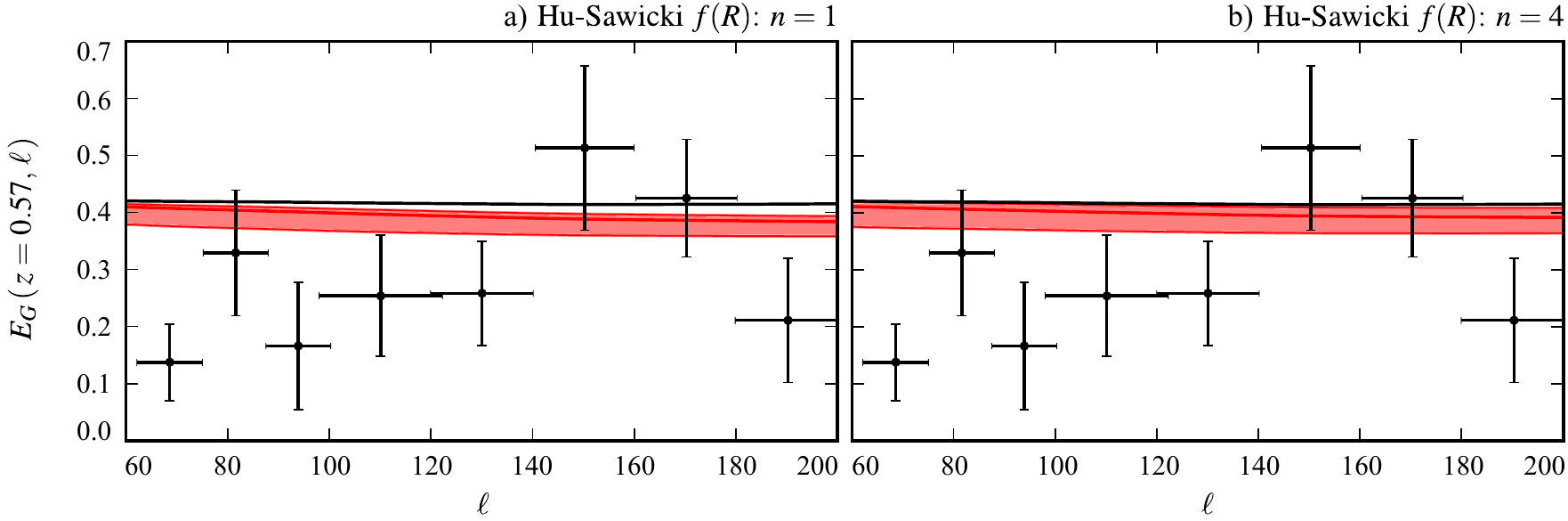}
\caption{
$E_G$ for Hu-Sawicki $f(R)$ gravity for a BOSS-CMASS-like survey with average redshift $\bar z=0.57$ and different angular scales. Different panels correspond to different values of the Hu-Sawicki exponent $n$. 
The black thick line corresponds to $\Lambda$CDM cosmology.
The red thick line shows the behavior of $E_G$  for the mean value of cosmological parameters of the corresponding model, as obtained from the most complete data set combination discussed in section~\ref{Sec:CosmologicalConstraints}.
The color bands denote the $1\sigma$ uncertainties on the $f^0_R$ parameter from the data compilation of Planck15+BAO+JLA+WiggleZ+CFHTLenS.
The black points with error bars measurements of $E_G$ from BOSS-CMASS~\protect\cite{Pullen:2015vtb}.
}
\label{fig:EG}
\end{figure}

After we implement the background solver, EFTCAMB solves consistently the full  perturbed Einstein-Boltzmann equations to compute all cosmological quantities of interest. We shall not review the matter and cosmic microwave background spectra but rather refer the reader to~\cite{Hu:2014oga}, where such observables have been outputted using EFTCAMB for designer $f(R)$ models. In this section we will focus instead on some specific combinations of observables that capture interesting phenomenological signatures, namely the growth rate $f\sigma_8$ and the $E_G$ statitstics, defined below. For some other interesting features see also~\cite{Bianchini:2015iaa}.

We shall first consider the observational effects imprinted in the time and scale dependence of $f\sigma_8$. 
Within General Relativity and the $\Lambda$CDM model, on linear scales, the combination of $f(z,k)={\rm d}\log \delta_m(z,k)/{\rm d}\log a$ and the amplitude of the linear power spectrum on the scale of $8\,h^{-1}\,\mbox{Mpc}$ does not depend on the scale at which this quantity is computed. This does not hold in generic modified gravity models.
In particular, in $f(R)$ models, the growth rate is enhanced at scales that are smaller than the Compton wavelength of the scalaron field, resulting in specific scale-dependent patterns \cite{Song:2006ej,Pogosian:2007sw}. 

In figure~\ref{fig:fs8} we show the effect of this scale dependence in $f\sigma_8$, at two different scales.
It can be noticed that  in Hu-Sawicki $f(R)$ gravity, the growth rate of short wavelength modes  is enhanced with respect to that of long wavelength modes and also as compared to the scale-independent rate of the $\Lambda$CDM model.
This behavior is consistent with the picture emerging from the matter power spectra, as those shown in~\cite{Hu:2013twa}. While being mild, this scale dependence cannot certainly be neglected and the plots in figure~\ref{fig:fs8} warn about the possibility of biasing cosmological inference by not properly accounting for it.
From figure~\ref{fig:fs8}, by comparing the two panels, we can also see how this scale dependence changes with the value of the parameter $n$. The smaller the  value of $n$, the stronger the enhancement of the growth with respect to $\Lambda$CDM and hence the bigger the overall scale dependence of $f\sigma_8$.

The second observable that we shall consider is the modified gravity statistic, $E_G$,~\cite{Zhang:2007nk} that was proposed to constrain or detect deviations from a $\Lambda$CDM cosmology.
In order to do so, different observables are combined into a single one that is free from the modeling of other unknown quantities like bias.
We refer to~\cite{Leonard:2015cba} for the complete derivation of the observationally-motivated definition of $E_G$  in modified gravity scenarios.
Here we consider the following definition, given by~\cite{Pullen:2014fva,Pullen:2015vtb}:
\begin{equation} \label{eq:EG}
E_G = \Gamma\frac{C_{\ell}^{\kappa g}}{C_{\ell}^{\theta g}},\qquad \Gamma = \frac{\mathcal{H}(z)f_g(z)}{3H_0^2 W_{\kappa}(z)(1+z)^2}\;,
\end{equation}
where $C_{\ell}^{\kappa g}$ is the cross correlation between galaxy lensing convergence and galaxy number counts fluctuations; $C_{\ell}^{\theta g}$ is the cross correlation between galaxy peculiar velocity and galaxy number counts fluctuations; $W_{\kappa}(z)$ is the lensing convergence window function; $f_g(z)$ is the normalized galaxy redshift distribution as in~\cite{Font-Ribera:2013rwa}. 

The particular combination of observables in~(\ref{eq:EG}),  is scale independent on linear scales for the $\Lambda$CDM model. 
In general, when we consider modified gravity scenario, this does not hold anymore and $E_G$ acquires a scale dependence.
This is shown for the Hu-Sawicki model in figure~\ref{fig:EG}. While the $\Lambda$CDM behavior is clearly scale independent the Hu-Sawicki case shows a slight scale dependent suppression.
This suppression is due to an enhancement of the galaxy velocity field in $f(R)$ gravity with respect to the $\Lambda$CDM case.
As in the previous case we can notice that the scale dependence and deviation from the $\Lambda$CDM behavior is stronger in the $n=1$ case and weaker in the $n=4$ case.

\clearpage

\section{Cosmological constraints}\label{Sec:CosmologicalConstraints}

\begin{table}
\centering
\begin{tabular}{|c|l|c|c|c|}
\hline
Model & Data set              & $\log_{10}(-f^0_{R})$      & $\sigma_8$ & $H_0$ \\
\hline
\multirow{7}{*}{$n=1$} 	& $D_1$           					&  $<-2.7$					& $0.87^{+0.12}_{-0.08}$		& $67.7^{+0.9}_{-0.9}$ \\
\\
					& $D_1+{\rm WiggleZ}$           		&  $-3.4^{+1.4}_{-1.2}$		& $0.95^{+0.07}_{-0.07}$		& $67.7^{+0.9}_{-0.9}$ \\
\\
					& $D_1+{\rm CFHTLenS}$           	&  $<-4.5$					& $0.83^{+0.07}_{-0.04}$		& $67.8^{+0.9}_{-0.9}$ \\
\\
					& ${\rm All}$           				&  $<-3.2$					& $0.90^{+0.07}_{-0.10}$		& $67.7^{+0.9}_{-0.9}$ \\
\hline
\multirow{7}{*}{$n=4$} 	& $D_1$           					&  $-$				& $0.87^{+0.06}_{-0.07}$		& $66^{+2}_{-4}$\\
\\
					& $D_1+{\rm WiggleZ}$           		&  $>-5.2$					& $0.90^{+0.04}_{-0.08}$		& $66^{+2}_{-4}$\\
\\
					& $D_1+{\rm CFHTLenS}$           	&  $<-2.8$					& $0.84^{+0.07}_{-0.04}$		& $67.8^{+0.9}_{-0.9}$\\
\\
					& ${\rm All}$           				&  $-4.4^{+2.9}_{-4.2}$		& $0.87^{+0.06}_{-0.07}$		& $67^{+1}_{-1}$\\
\hline
\multirow{7}{*}{$n$~free} 	& $D_1$           					&  $-$				& $0.86^{+0.08}_{-0.07}$		& $67^{+2}_{-3}$\\
\\
					& $D_1+{\rm WiggleZ}$           		&  $>-5.2$					& $0.93^{+0.09}_{-0.11}$		& $67^{+2}_{-2}$\\
\\
					& $D_1+{\rm CFHTLenS}$           	&  $<-2.3$					& $0.84^{+0.06}_{-0.04}$		& $67^{+1}_{-1}$\\
\\
					& ${\rm All}$           				&  $<-1.9$					& $0.88^{+0.08}_{-0.08}$		& $67^{+1}_{-1}$\\
\hline
\end{tabular}
\caption{\label{Tab:CP}
The $95\%$ C.L. marginalized bounds, from the used data set combinations, on Hu-Sawicki models.
In the case where the parameter $n$ is allowed to vary, no significant constraints on $n$ is found.}
\end{table}

We shall now use different cosmological data to place constrain on the Hu-Sawicki model via EFTCAMB/EFTCosmoMC. Cosmological constraints on this model have been already explored in~\cite{Bean:2006up,Lombriser:2010mp,Lombriser:2011zw,Lombriser:2012nn,Lombriser:2013wta,Terukina:2013eqa,Hu:2012td,Xu:2015usa,Hojjati:2015ojt} (see \cite{DeFelice:2010aj,Koyama:2015vza} for review). Let us stress that most of the previous approaches have relied on the quasi-static approximation and often assumed a $\Lambda$CDM background, while we evolve the full dynamics of perturbations and use the specific background expansion history of the model. In our analysis  we  use several geometrical and dynamical probes, combining them progressively. 

The baseline data set combination employed, hereafter ``D1'', consists of: the {\it Planck} full mission $29$ months temperature and polarization low multipole likelihood ($\ell\leq29$,~\texttt{lowTEB})~\cite{Aghanim:2015xee}; 
the {\it Planck} CMB temperature and polarization likelihood (TT+TE+EE,~\texttt{Plik})~\cite{Aghanim:2015xee} for the high-$\ell$ modes ($30\leq\ell\leq2508$) from the $100$, $143$, and $217$ GHz frequency channels; 
the CMB lensing likelihood ($40\leq\ell\leq400$) from {\it Planck}-2015~\cite{Ade:2015zua} that resulted in a $40\sigma$ detection of the lensing signal;
the ``Joint Light-curve Analysis'' (JLA) Supernovae sample as analysed in~\cite{Betoule:2014frx} which is constructed from the SNLS, SDSS and HST SNe data, together with several low redshift SNe;
baryon acoustic oscillations measurements taken from the SDSS Main Galaxy Sample at $z_{\rm eff}=0.15$~\cite{Ross:2014qpa},  the BOSS DR11 ``LOWZ" sample at $z_{\rm eff}=0.32$~\cite{Anderson:2013zyy}, the BOSS DR11 CMASS at $z_{\rm eff}=0.57$ of~\cite{Anderson:2013zyy} and the 6dFGS survey at $z_{\rm eff}=0.106$~\cite{Beutler:2011hx}.

We shall add to the D1 data set compilation the galaxy number density power spectrum from the WiggleZ Dark Energy Survey~\footnote{\url{http://smp.uq.edu.au/wigglez-data}} in order to exploit the constraining power of the galaxy clutering data. The WiggleZ data set consists of the galaxy power spectrum measured from spectroscopic redshifts of $170,352$ blue emission line galaxies over a volume of $1\,\mbox{Gpc}^3$~\cite{Drinkwater:2009sd,Parkinson:2012vd}. It has been shown by~\cite{Parkinson:2012vd,Blake:2010xz,Dossett:2014oia} that linear theory predictions are a good fit to the data regardless of non-linear corrections up to a scale of $k\sim 0.2\,\mbox{h}/\mbox{Mpc}$. Even though the modification of the growth rate could slightly alter this scale, in this work, we adopt a very conservative strategy and use the WiggleZ galaxy power spectrum with a cut off at $k_{\rm max} = 0.1\,\mbox{h}/\mbox{Mpc}$. We marginalize analytically over the standard scale independent linear galaxy bias for each of the four redshift bins, as in.~\cite{Parkinson:2012vd}. Hereafter, we shall refer to the combination of the D1 data set to the WiggleZ data set as ``D1+WiggleZ''. 

The third data set combination, that we dub ``D1+CFHTLenS'', consists of D1 joined with the measurements of the galaxy weak lensing shear correlation function as provided by the Canada-France-Hawaii Telescope Lensing Survey (CFHTLenS)~\cite{Heymans:2013fya}. 
This is a $154$ square degree multi-color survey that spans redshifts ranging from $z\sim0.2$ to $z\sim 1.3$ and is optimized for weak lensing analyses.
Here we consider the $6$ redshift bins tomography and we applied ultra-conservative cuts, as in~\cite{Ade:2015rim}, that exclude $\xi_{-}$ completely and cut the $\xi_+$ measurements at scales smaller than $\theta=17'$ for all the tomographic redshift bins.
These cuts make the CFHTLenS data insensitive to the modeling of the non-linear evolution of the power spectrum, as discussed in~\cite{Ade:2015rim}.

Finally we shall refer to ``All'' to indicate the data set combination obtained by joining D1, WiggleZ and CFHTLenS. We shall consider three different Hu-Sawicki models: a model with the exponent $n=1$, a model with the exponent $n=4$ and a model in which $n$ is free to vary in the range $[0,10]$. In all the cases, there is another model parameter which is left free to vary, i.e. $f_R^0$. We sample it logarithmically within the range $\log_{10}(-f^0_R)\in(-9,0)$. 	

The marginalized bounds on the model parameters, that we obtain for all cases are summarized in table~\ref{Tab:CP}. One can notice that bounds on $\log_{10}(-f^0_{R})$  from the D1 data set are weak for the $n=1$ model, and do not result in statistically significant constraints in the $n=4$ and $n$ free cases.
On the other hand the WiggleZ data set clearly favors a high value of $\log_{10}(-f^0_{R})$  driving the parameter bound away from the $\Lambda$CDM limit as far as:
\begin{align}
\log_{10}(-f^0_{R})=-3.4^{+1.4}_{-1.2}\;\;,~{\rm at}~95\%~{\rm C.L.}~(n=1,~{\rm  ``D1+WiggleZ''})
\end{align} 
A similar result was found, for another type of $f(R)$ model, in~\cite{Dossett:2014oia,Hu:2014sea}.
The CFHTLenS data set on the contrary pushes the bounds on $\log_{10}(-f^0_{R})$ toward the $\Lambda$CDM limit.
This behavior is caused by the fact that, in $f(R)$ gravity, the growth of cosmic structures, on scales below the Compton wavelength of the scalar field, is enhanced with respect to the $\Lambda$CDM scenario.
The combination of {\it Planck} and WiggleZ data seems to prefer this enhancement of the growth, while the CFHTLenS weak lensing data, favors the opposite behavior, {\it i.e.} a suppression of the amplitude of the matter power spectrum.  This is also reflected in the $\sigma_8$ bounds. If we consider the $n=1$ case we can immediately read from table~\ref{Tab:CP} that the combination of D1 with WiggleZ results in a statistically significant enhancement of the $\sigma_8$ value while the opposite holds for D1 joined with the CFHTLenS data set.
Furthermore, we can also see that the estimated Hubble parameter $H_0$ is weakly dependent on the data set and model considered. However, even if the mean value does not change significantly, the 95\% C.L. bounds are, instead, significantly dependent on the model considered. 

\begin{figure}
\centering
\includegraphics[width=\textwidth]{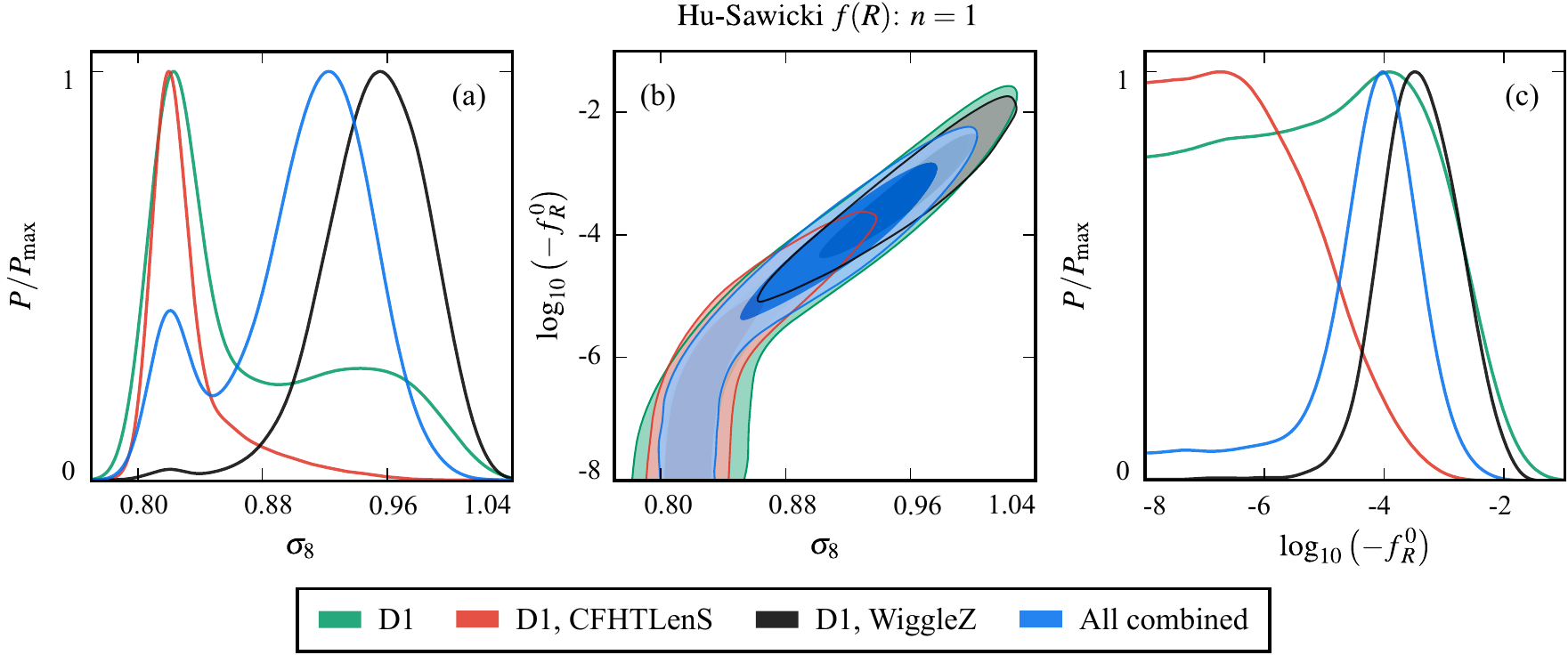}
\caption{Results of the analysis of Hu-Sawicki $f(R)$ with $n=1$.
{\it Panel (a):} The marginalized posterior of $\sigma_8$.
{\it Panel (b):} The marginalized joint posterior of $\sigma_8$ and $\log_{10}\left( -f^0_{R}\right)$.
The darker and lighter shades correspond respectively to the 68\% C.L. and the 95\% C.L..
We can see the strong degeneracy between these two parameters.  The WiggleZ data set drives the joint posterior toward high values  of $\log_{10}\left( -f^0_{R}\right)$ and $\sigma_8$.
{\it Panel (c):} The marginalized posterior of $\log_{10}\left( -f^0_{R}\right)$. As we can see the WiggleZ data set favors a large value of $\log_{10}\left( -f^0_{R}\right)$.
In all panels different colors correspond to different data set combinations as shown in legend.}
\label{fig:chains_n1}
\end{figure}

\begin{figure}
\centering
\includegraphics[width=\textwidth]{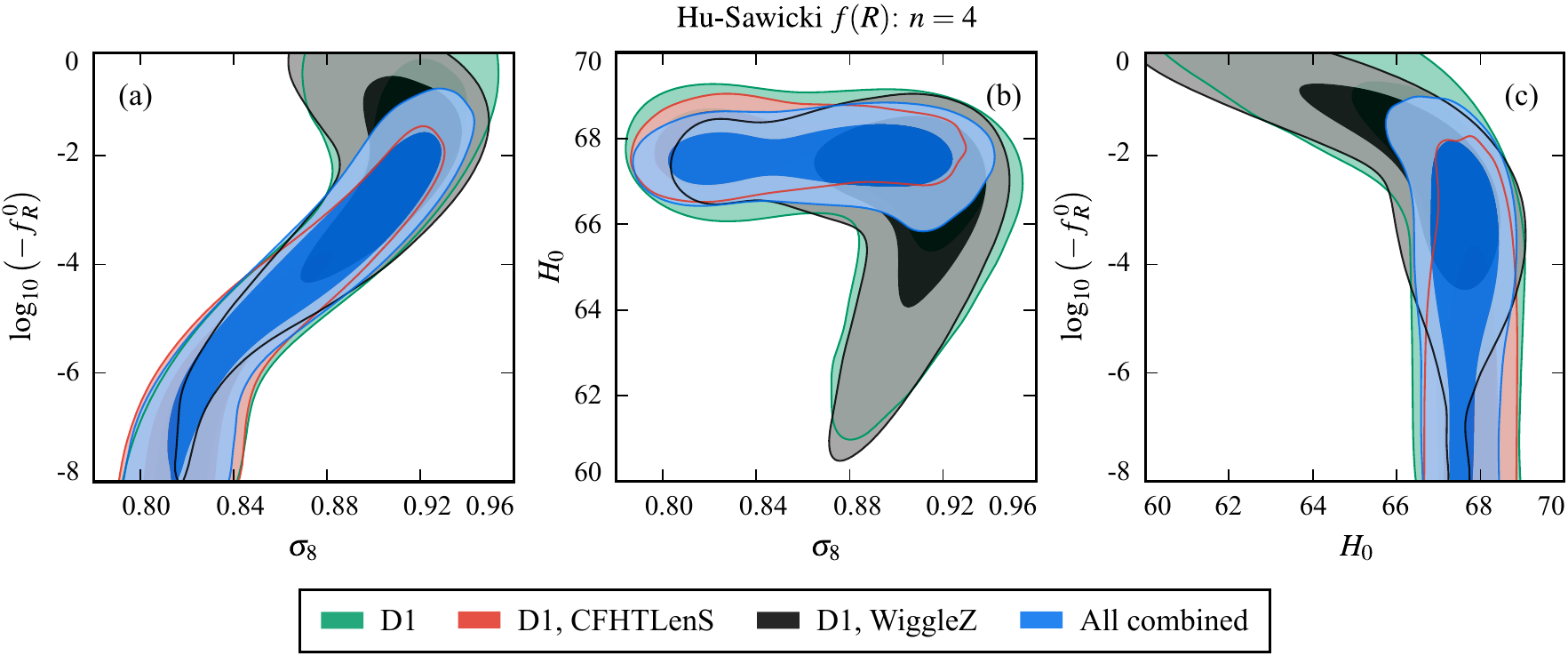}
\caption{Results of the analysis of Hu-Sawicki $f(R)$ with $n=4$.
{\it Panel (a):} The marginalized joint posterior of $\log_{10}\left( -f^0_{R}\right)$ and $\sigma_8$. We can see the well known degeneracy between $\log_{10}\left( -f^0_{R}\right)$ and the growth of structure.
{\it Panel (b):} The marginalized joint posterior of $H_0$ and $\sigma_8$ displaying a marked degeneracy between the two parameters.
{\it Panel (c):} The marginalized joint posterior of $\log_{10}\left( -f^0_{R}\right)$ and $H_0$.
We can see a strong degeneracy between the two parameters for large values of $\log_{10}\left( -f^0_{R}\right)$. 
This is introduced by the modification of the background equations.
In all panels different colors correspond to different data set combinations as shown in legend.}
\label{fig:chains_n4}
\end{figure}

\begin{figure}
\centering
\includegraphics[width=\textwidth]{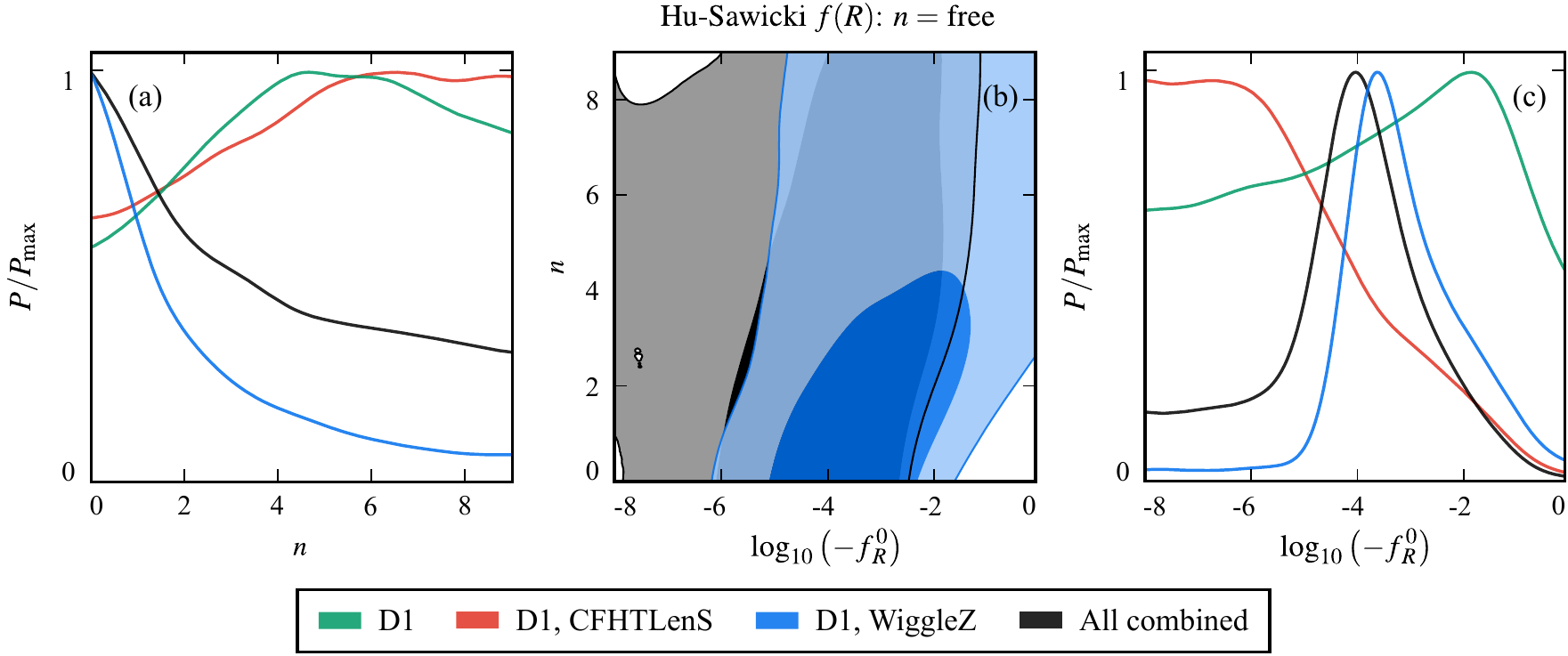}
\caption{Results of the analysis of Hu-Sawicki $f(R)$ varying both model parameters, i.e. the index $n$ and the boundary condition $f_R^0$. 
{\it Panel (a):} The marginalized posterior of $n$. As we can see there is no statistically significant bound on this parameter. 
{\it Panel (b):} The marginalized joint posterior of $n$ and $\log_{10}\left( -f^0_{R}\right)$.
The darker and lighter shades correspond respectively to the 68\% C.L. and the 95\% C.L.. 
It is the first time were the significant correlation between these two parameters is shown.
{\it Panel (c):} The marginalized posterior of $\log_{10}\left( -f^0_{R}\right)$.
As in the previous cases the WiggleZ data set drives the posterior of $\log_{10}\left( -f^0_{R}\right)$ away from the $\Lambda$CDM limit.
In all panels different colors correspond to different data set combinations as shown in legend.}
\label{fig:chains_nf}
\end{figure}

All these conclusions about the parameters bounds are further confirmed by the inspection of the marginalized posterior of several parameters. In figure~\ref{fig:chains_n1} we can clearly see the degeneracy between the present day amplitude of scalar perturbations, $\sigma_8$, and the present day value of $f_R$. 
Through this degeneracy,  WiggleZ data set favors a high value of both parameters, while CMB and CFHTLenS measurements favor a smaller value. This degeneracy changes as we change the value of the Hu-Sawicki model exponent. 
With this respect, figure~\ref{fig:chains_n4}, shows the $n=4$ case.
Noticeably, we can see that as soon as  $\log_{10}(-f^0_{R})>-2$, the degeneracy between $\sigma_8$ and $\log_{10}(-f^0_{R})$ has an abrupt change in direction.
This change in direction clearly shows up also in the posterior distribution of $\sigma_8$ and $H_0$, in panel (b) of figure~\ref{fig:chains_n4}.
The reason for this behavior can be understood by looking at panel (c) of the same figure.
As we can clearly see, for $\log_{10}(-f^0_{R})<-2$, the parameter describing $f(R)$ is not degenerate with the Hubble parameter but, as soon as $\log_{10}(-f^0_{R})>-2$, a marked degeneracy arises. 
This means that when $\log_{10}(-f^0_{R})<-2$ the model is constrained through its effect on perturbations while in the regime $\log_{10}(-f^0_{R})>-2$ the effect of this modification of gravity at the background level is not negligible. On the contrary, in this parameter range, background observables play an important role in constraining the model.
To further support this conclusion we notice that such a degeneracy does not arise in designer $f(R)$ models, where the cosmological background mimics exactly the $\Lambda$CDM one, as in~\cite{Hu:2014sea}. When $n$ is allowed to vary the situation slightly changes. As we can see from figure~\ref{fig:chains_nf} there is a significant degeneracy between $n$ and $\log_{10}(-f^0_{R})$ and, being $n$ weakly constrained, the bound on $\log_{10}(-f^0_{R})$ gets weaker. 

\begin{figure}
\centering
\includegraphics[width=\textwidth]{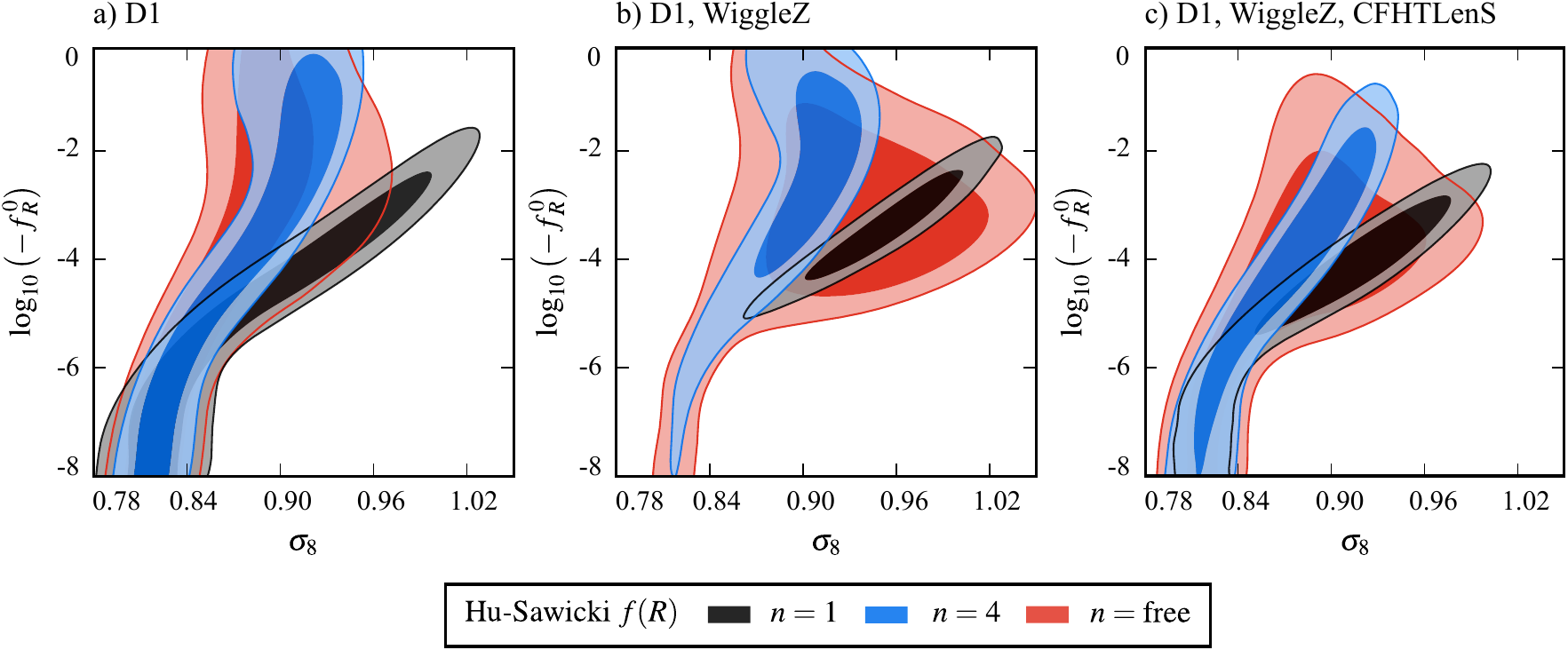}
\caption{All panels show the marginalized joint posterior of $\sigma_8$ and $\log_{10}\left( -f^0_{R}\right)$ for different data set combinations.
The darker and lighter shades correspond respectively to the 68\% C.L. and the 95\% C.L.. Different colors correspond to different Hu-Sawicki models as shown in legend.
We can see the change in the degeneracy direction depending on the value of $n$ that is used, with a weak dependence on the data set.
}
\label{fig:chains_mod}
\end{figure}

Finally, in figure~\ref{fig:chains_mod}, for different data set combinations we can see how the degeneracy between the growth of structure and modifications of gravity induced by the Hu-Sawicki model depends on the value of the exponent $n$.
The general trend is that the degeneracy is stronger for small values of $n$ and weaker for high values on $n$, as can be seen clearly from the $n=1$ and $n=4$ posterior. 
The case with $n$ free obviously covers these two sub-cases and noticeably enough, in this case, the degeneracy is not that different from the other two cases.

\clearpage

\clearpage

\section{Conclusions}\label{conclusion}

In this paper, we have presented in detail the procedure at the basis of the full mapping of a specific model of modified gravity into the linear Einstein-Boltzmann solver EFTCAMB. 
We chose the Hu-Sawicki $f(R)$ model as an example, demonstrating how to set up its background solver and how to implement it in EFTCAMB, mapping between the model parameters and the relevant EFT functions. 
Once the mapping of the model into the EFT language is worked out, from the numerical point of view all  the user  needs for this model is to interface EFTCAMB with the background equation of state of the dynamical dark energy field (or the modified Hubble parameter) and the EFT functions. 
The relevant modules of the code will automatically calculate their derivatives and integrals numerically.  More generally, one will need to interface EFTCAMB with the time evolution of the EFT functions corresponding to the given model. The advantage of this mapping algorithm is that it allows the users to implement a specific model within a few steps without going to the details of the complicated perturbation equations. The EFTCAMB background and perturbation solvers will solve the coupled system consistently.

In order to display the full potential of this implementation, which treats fully and in a model specific way the dynamics of both the background and the perturbations, we discussed some linear structure growth rate estimators, such as $f\sigma_8$ and $E_G$, and studied the constraints on the Hu-Sawicki $f(R)$ model with current cosmological data. 
As for the growth estimators, unlike in the $\Lambda$CDM scenario, at the linear regime, both $f\sigma_8$ and $E_G$ are \emph{scale-dependent} in DE/MG models, and this was clearly visible for the Hu-Sawicki model. For this reason, we showed the $f\sigma_8$ growth in different wavenumber and the $E_G$ angular distribution at some fixed averaged redshift snapshot.  The former, as expected, showed quite significant scale-dependent profiles against the present redshift space distortion data. However,  since there are a lot of residual systematics in the $E_G$ estimator pipeline, the scale-dependent angular distribution in $f(R)$ case is still inside of the scattering of the current data points.

Finally, we run a Markov-Chain Monte-Carlo analysis and estimated the model and cosmological parameters against Planck CMB, including CMB lensing, WiggleZ galaxy number density counts as well CFHTLenS weak lensing surveys. We found some degeneracy between $\sigma_8$ and $|f^0_R|$, through which, the WiggleZ data set, favors a high value of both parameters, while CMB and CFHTLenS measurements favor a smaller value. Furthermore, when $n=4$ and  $\log_{10}(-f^0_{R})>-2$, this degeneracy has an abrupt change in direction.
This change in the degeneracy direction clearly showed up also in the posterior distribution of $\sigma_8$ and $H_0$. For the small value of the scalaron Compton wavelength, such as $\log_{10}(-f^0_{R})<-2$, the parameter describing $f(R)$ is not degenerate with the Hubble parameter but, as soon as $\log_{10}(-f^0_{R})>-2$, a marked degeneracy arises. It is because when $\log_{10}(-f^0_{R})<-2$ the model is constrained through its effect on perturbations, while in the regime $\log_{10}(-f^0_{R})>-2$ the effect of this modification of gravity at the background level is not negligible. Hence, the background kinematics play an important role in constraining the model in this parameter range. This degeneracy was fully displayed for the first time since with our procedure we have solved the specific background for the model, without approximating it to a $\Lambda$CDM one.

\section*{Acknowledgements}
We are grateful to Nicola Bartolo, Antonio J. Cuesta, Wayne Hu, Michele Liguori and Massimo Viola for useful discussions and comments on the manuscript.
BH is partially supported by the Dutch Foundation for Fundamental Research on Matter (FOM) and Programa Beatriu de Pin\'os. Marco Raveri acknowledges partial support from the INFN-INDARK initiative. Marco Raveri and Matteo Rizzato acknowledge the Instituut Lorentz (Leiden University) for hospitality during the time when this research was carried out. AS acknowledges support from The Netherlands Organization for Scientific Research (NWO/OCW), and also from the D-ITP consortium, a program of the Netherlands Organisation for Scientific Research (NWO) that is funded by the Dutch Ministry of Education, Culture and Science (OCW).




\bibliographystyle{mnras}

\bibliography{mnras_hs}







\bsp	
\label{lastpage}
\end{document}